\documentclass[pre,longbibliography]{revtex4-1}
\usepackage{graphicx}
\usepackage[utf8]{inputenc}
\usepackage[usenames,dvipsnames,svgnames,table]{xcolor}
\usepackage{amsmath,amssymb,latexsym,graphics,epsfig}
\usepackage{pstricks}
\usepackage{cancel}
\usepackage{color}
\usepackage{graphicx}
\usepackage{natbib}
\DeclareGraphicsExtensions{.pdf,.eps,.png}
\begin{document}\title{Kinetics of coarsening  have dramatic effects on the
microstructure: self-similarity  breakdown induced by viscosity contrast. }
\author{Herv\'e Henry$^1$ and Gy\"orgy Tegze$^2$}
\affiliation{$^1$
 Laboratoire de Physique de la Matière Condensée, École Polytechnique, CNRS, Université Paris-Saclay, 91128 Palaiseau  Cedex, France
}
\affiliation{$^2$Wigner Research Centre for Physics, P.O. Box 49, H-1525 Budapest, Hungary}
\date{\today}
\begin{abstract}
The viscous coarsening of a phase separated mixture is studied and the effects
of the viscosity contrast between the phases are investigated. 
From an analysis
of the microstructure, it appears that for moderate departure from the
perfectly symmetric regime the self-similar bicontinuous  regime is robust. However, the
connectivity of one phase decreases when its volume fraction  decreases or when
it is becoming less viscous than the complementary phase.
Eventually self-similarity breakdown is observed and characterized.
\end{abstract}
\maketitle

\section{Introduction}
 The phase separation and the subsequent coarsening of the microstructure  under the
 effects of the surface tension is an  ubiquitous mechanism in industrial
 processes\cite{Levitz91,Craievich1986,Kumar_1996_PRL}. In this context understanding how patterns are formed and how they
 can be controlled is highly desirable. However this process is complex and
 involves different mechanisms. Indeed, after a quench an initially
 thermodynamically stable mixture will lose stability\cite{Cahn1958}, it will then phase
 separate spontaneously through spinodal decomposition\cite{Cahn1965}.
 In both cases, after the initial phase separation process, a complex
 microstructure has spontaneously formed. It is constituted of the two phases
 that are separated by an interface with a huge surface area. The evolution of
 the system will then be driven by the surface tension and will lead to an
 increase of the characteristic lengthscale of the pattern $l$. If both phases
 are liquid, the coarsening process involves  two successive regimes. First, when $l$ is
 small the coarsening is mostly due to diffusion and $l$ grows as
 $t^{1/3}$\cite{Cahn1966diffusive,LSW61,Kwon2007,Kwon2010,Puri1997,Sun2018},
 thereafter, when $l$ is large the effects of fluid flow become
 dominant\cite{expe2,Siggia1979,Appert1995visc,Bastea97visc,Kendon2001inertial,Tanaka1998} and
 $l$ grows as $t$. Hence the time evolution of the characteristic lengthscale
 of the microstructure is well understood. However, the understanding of the
 microstructure itself and of how it can be controlled is limited. Indeed, the
 volume fraction of the phases can be used to control the microstructure and for
 instance by
 properly choosing volume fractions of the phases, one can tune a transition
 from a bicontinuous microstructures where both phases are percolating clusters
 to an inclusions in a matrix pattern.

  This approach is mainly focused on
 the initial phase separation process and overlooks the importance of the
 kinetics of the coarsening process, which is  well exemplified by recent
 experiments on the viscous coarsening of glasses\cite{Bouttes2014,Bouttes2016}. 
 Indeed  it has been  shown that the
 microstructure is affected by both the volume fraction of the phases and their
 relative viscosities. Hence, while the driving force for coarsening is always  the
 reduction of the surface energy, the kinetic of the coarsening, that is the
 path taken by the system to dissipate energy,  has a
 dramatic effect on the microstructure. For instance if the volume fraction of
 the minority phase is close to $0.3$, when it is much more viscous than the
 majority phase, a bicontinuous microstructure  remains during coarsening
 while if it is much less viscous a transition toward a discontinuous
 microstructure is observed. 
 This  is in line with previous theoretical work where 
 the interplay between diffusion and flow during the initial stage of spinodal
 composition was studied\cite{Tanaka1998} or where visco-elastic effects were 
 taken into account\cite{Tanaka2000,Araki2000,Kumar_1996_PRL}

 Here, we focus on the late stage of coarsening in liquids where the pattern
 evolution is due to viscous flow  (diffusion can be neglected) and where the
 inertial effects are also negligible. 
 Using numerical simulations  we show that tuning the kinetic of the
 coarsening process through the viscosity ratio between the phases  
 dramatically changes  the microstructure. The paper is
 organised as follows. First we present the model equation, the numerical
 methods and the choice of the initial conditions and of parameters in the light
 of the physics  of the coarsening process. We also describe briefly the 
 tools that have been used to describe the microstructure. Thereafter we present
 the numerical results in the case where the self-similar coarsening is robust
 and we discuss its loss of stability. Finally we conclude. 

\begin{figure}
  \centerline{\includegraphics[width=0.6\textwidth]{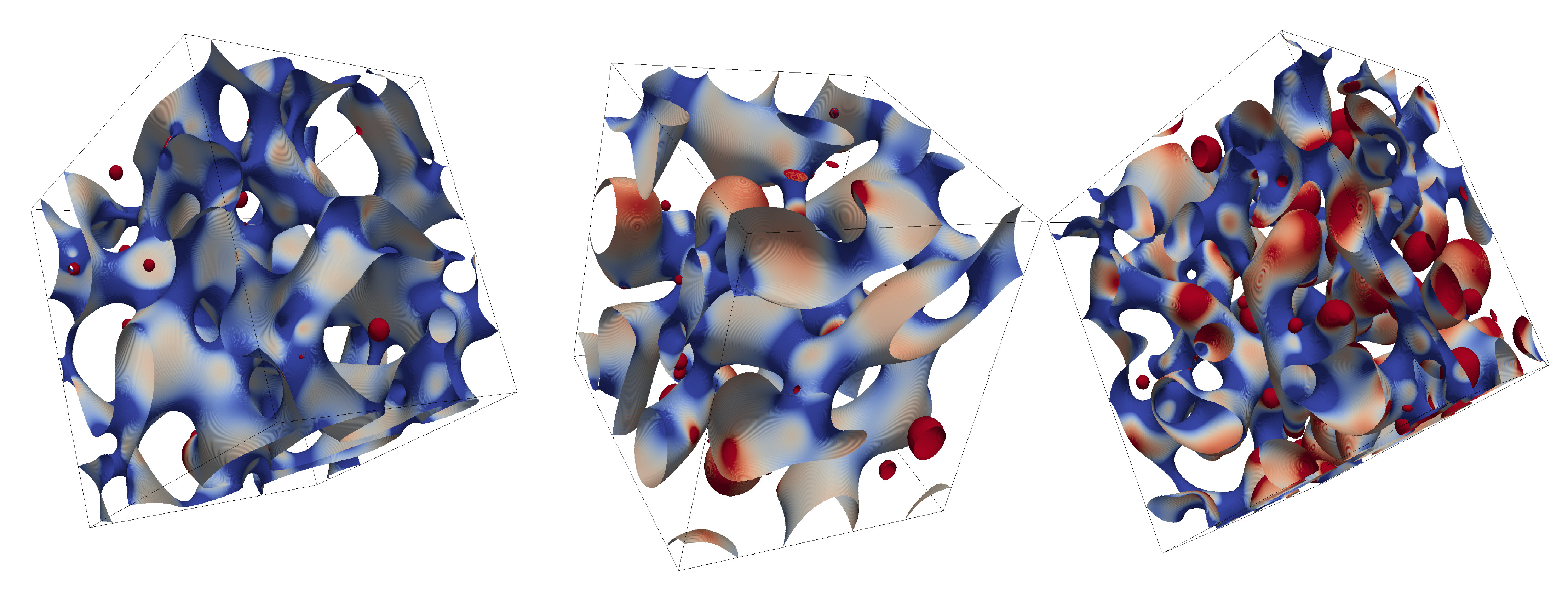}}
  \caption{Typical images of the interface between the phases for different
  values of the parameters. The color of the surface indicates the gaussian
  curvature (blue: negative, red: positive). One can see that the regions with
  positive gaussian curvature are more present in the later case.
  \label{snapshots_VC}}
\end{figure}

 \section{Method}

{The thermodynamics of a binary fluid is well described by the diffuse interface theory of Cahn and Hilliard\cite{Cahn1958}.
The simplest symmetric form of the Cahn-Hilliard free-energy reads as:
\begin{equation}
  \mathcal F = \int \epsilon^2 (\nabla c)^2 + A (c^2(c-1)^2)
\end{equation}
}
 With such a choice, when
$A>0$, an homogeneous mixture with a composition close to 0.5 will spontaneously
phase separate into two phases with concentration 0 and 1. The surface tension
associated to the interface between the phases and its thickness  can be chosen
by adjusting $A$ and $\epsilon$. Here $\epsilon^2=2.56$ and $A=4.$  were chosen so that
$\gamma\approx0.75$\cite{Kendon2001inertial} and so that the interface thickness is
of the order of $w_{int}=1.6$\cite{Cahn1958}.  The coarsening dynamics via convection and
diffusion is governed by the coupled Navier-Stokes  (eq.\ref{eq:NS2}) and the
convective Cahn-Hilliard (CH) (eq.\ref{eq:CH}) equations (NSCH), also known as
model H \cite{Hohenberg_RMP}. The Navier-Stokes/Cahn-Hilliard\cite{Anderson1997}
model was used along with the incompressibility constraint (eq.\ref{eq:PE}):
\begin{eqnarray}
	\partial_t c + \mathbf{v} \cdot \nabla c &=& -M \triangle \mu \label{eq:CH}\\
	\partial_t \mathbf{v} + \nabla \cdot (\mathbf{v} \otimes \mathbf{v}) &=& { \frac{-1}{\rho}(\nabla p + c \nabla\mu)} \label{eq:NS1}\\
	&+&  \nabla \cdot \left(\frac{\nu{(c)}}{2}(\nabla\mathbf{v}+\nabla\mathbf{v}^T)\right)\label{eq:NS2} \nonumber\\
\nabla\cdot \mathbf{v} &=& 0 \label{eq:PE}
\end{eqnarray}
In the Cahn-Hilliard equation (Eq. \ref{eq:CH}),  $M$ is the mobility,
$\mu=\delta \mathcal F / \delta c$ is the chemical potential that derives from
the CH free energy. 

In the Navier-Stokes equation (Eq. \ref{eq:NS1})
the  $-\nabla p$ term on the RHS includes a Lagrangian multiplier that forces incompressibility.
The second  term  is the thermodynamic stress,  and accounts for capillary forces.  
The last term accounts for the viscous dissipation with a composition dependent kinematic  viscosity:
\begin{equation}\nu{(c)}=(1-c)\nu_1+c\ \nu_2,
\end{equation}
where $\nu_1$  and $\nu_2$ are the viscosities of the two phases corresponding to $c=0$ and $c=1$. The viscosity contrast is then defined as $VC=\nu_1/\nu_2$
The mass density is $\rho$. The model equations were simulated numerically  using standard
approaches\cite{Orszag_1969_PHYSFLUIDS,Orszag_1972_PRL,Liu_2003_PHYSD,Zhu_1999_PRE}
that are described in the supplementary material  of\cite{henrytegze2017} together
with a more detailed  description of the model
equations that is inspired by \cite{Barry2011,Gyula2016PRE}. 

The NSCH model reproduces well the initial phase separation followed by the
coarsening of the microstructure that is due to diffusion at small length-scales
with a characteristic length-scale growing as
$l\propto t^{1/3}$\cite{LSW61,Cahn1966diffusive}. At larger length scales the coarsening
is driven by convection that is governed by surface tension and viscous
dissipation. As a result $l$ grows linearly: $l= v_0t \propto \gamma/\nu t$
\cite{Siggia1979} where $\gamma$ is the surface tension and $\nu$ is the
viscosity of the fluid when $VC=1$.
The transition from the diffusive growth to a viscous growth occurs  when $v_0$ is much larger than the growth velocity associated 
with diffusion (which itself  is a function of the mobility of chemical
species { and of the chemical potential difference induced by the Gibbs effect}). This translates into the fact that the Péclet number
($Pe=lv_0/M/\gamma$)  is large. 
Finally the viscous growth law  looses its validity   when  inertial effects
cannot  be neglected (the Reynolds number $Re$, defined as  $l/l_0$ where
$l_0=\nu^2/(\gamma\rho)$ becomes large). { Since, we are
considering fluids with different viscosities, two Reynolds number can be
computed: one for each phase. Since the fluid flow in both phases share the same
velocity  and since there is a clear relation between the fluid flow velocity
and the effective viscosity  $\nu_{eff}=\sqrt{\nu_0\nu_1}$\cite{henrytegze2017},
the Reynolds number in each  phases writes $Re_{1,2}=l/(\nu_{eff}^2\gamma \rho
VC^{\pm 1/2})$.}

Here we have limited ourselves to the
viscous coarsening of an already  phase separated mixture, assuming that the viscosity is
sufficiently high to avoid the effects of fluid flow  during the 
initial  phase separation and before well defined phases are present and 
coarsening takes place\cite{Tanaka1998}. It is important to note that during
the course of the coarsening, since $l$ is growing, these two numbers grow
(proportionally to $l$). As a
result, during the coarsening of a bicontinuous structure, both $Pe$ and $Re$
will increase and there is a transition from a diffusive coarsening regime
where ($Pe<<1$, $Re<<1$) to a viscous dominated regime ($Pe>>1$ and $Re<<1$) followed by an
inertia dominated regime ($Pe<<1$, $Re>>1$)\cite{Kendon1999scaling}. Here we have focused
on the well defined Siggia regime for which   $Pe>>1$ and $Re<<1$.   
These
constraints apply on the  macroscopic lengthscale. In contrast with this
requirement, at the scale of the interface, the  flow   deforms the concentration
profile through the interface and therefore  changes  the surface tension.
This effect is unwanted and needs to be counterbalanced by an appropriate
restoring mechanism. In actual systems this mechanism is diffusion which is
effective on the scale of the actual interface thickness. Here,  in order to allow computations, the interface thickness
is increased, and  some care must be taken to ensure that the diffusion is still
efficient enough to restore the equilibrium profile. This translates into the
fact that the interface Péclet number must be small enough.  According to  these
constraints and using the results of \cite{henrytegze2017}
we have chosen $\rho=1$, $\nu_{eff}=\sqrt{\nu_0\nu_1}=8$ and a mobility of
$M=0.0625$. 
With this choice of parameters and { viscosity contrasts ranging from 1 to 128}
both the inertial and  diffusive effects can be neglected during
coarsening for system characteristic length ranging from $\approx 20$ to $200$. 
 { However,  when considering viscosity contrast significantly  higher than 128, the decrease of the viscosity of the fluid phase is likely 
to lead to significant  a departure from the ideal low Re regime with parameters used here. In the case of experimental systems presented in \cite{Bouttes2014}, 
the viscosities of both phases are such that they are both in the low $Re$ regime. }

In our simulations the grid spacing is  set  to 1 and time step $\Delta t$ is also set  to
1. Typical domain size is $1024^3$ which ensures that finite size effects are
negligible.
\begin{figure}
  \centerline{\includegraphics[width=0.6\textwidth]{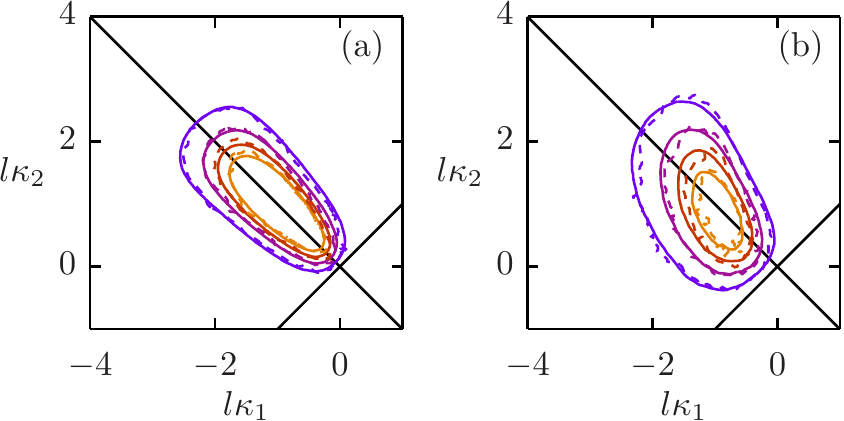}}
  \caption{\label{fig50}. Contour plot of the PDFs of the principal curvature
  for a volume fraction of $\varphi=0.5$ for two value of the viscosity
  contrast $VC=1$ (a) and $VC=128$ (b). The contour lines  are equally spaced.Taken
  from\cite{henrytegze2017}.Contours have been taken at two distinct times
  corresponding to (solid) $l \approx 23$ in (a)  ( resp. 30 in (b)) and to (dashed)
$l \approx 88$ in (a) (resp.84 in (b) ) (dashed).}
\end{figure}
 The constraints on the macroscopic Péclet number and on the interface  Péclet
 number imply that the initial phase separation is always  affected by the
 flow and by the viscosity contrast between the two phases. Therefore, we have
 chosen to use initial conditions that were computed through a well defined
 procedure (described in appendix). This approach allows us to focus on the effects of the
 coarsening process itself and had the advantage to allow to build different
 microstructures with different statistical properties in order to test the
 robustness of the self similar regime by showing whether the same self similar
 regime is reached from two different initial conditions that are not simply
 two realizations of the same stochastic process. During the build up of the
 initial condition the volume fraction of the phase 1,  $\varphi$ is set.

 Finally we present briefly the tools of analysis that were used here.
As in our previous work\cite{henrytegze2017}, the microstructure is first characterized by a characteristic lengthscale $l$ that is computed as the  ratio between the total volume and the
total interface between the phases. More precisely, it is defined using an energetic approach:
\begin{equation}
l=\frac{V \gamma}{\int \epsilon^2 (\nabla c)^2  }
\end{equation}
which gives actually $l=V/S$ when the interface between the phases corresponds
to an equilibrium profile. In order to characterize more finely the
microstructure  other quantities are studied. The geometry of the pattern is
described using the statistical properties of the curvature of the interface
between the phases. From the field of the implicitly defined interface, the
curvature are computed using implicit formulas\cite{Goldman2005}. 
The  Probability Distribution functions of the principal curvatures are then
determined as in\cite{henrytegze2017,Kwon2007,Kwon2010}. A typical example of the
PDFs contour that will be used as a \textit{reference} in the following  is shown in fig. \ref{fig50}. In addition integral
quantities are considered: the averaged mean curvature rescaled by $l$ and the
rescaled  genius number: $g=l^2(1-\int_V\kappa_g/(4\pi))/S$ where
$\kappa_g=\kappa_1\kappa_2$ is the  Gaussian curvature of the interface. 
The Gaussian curvature is independent of the orientation of the interface and the genius number
$g_l=(1-\int_V\kappa_g/(4\pi))$  is a topological invariant of the pattern which
is directly related to its Euler's characteristic. The orientation of the
surface is chosen so that  the normal points toward the inside of the phase for
which the volume fraction is given.

In addition, the
conductance $\mathcal{G}$  of the microstructure under the assumption that one phase is
conducting and the other is isolating (details of the computation and of the
numerical method are given in appendix) is also computed.
This  quantity gives an estimate of the connectivity of the phase. Indeed, when the phase is non  percolating it goes to zero while when it is percolating it can be viewed as the averaged total surface area  of the channels that go from one side of the system to the other.

\section{Results}
During the course of our simulations two distinct regimes have been observed. The
first one is a continuation of  the regime previously described
\cite{henrytegze2017} for the case where the volume fraction is 0.5. { However, 
considering volume fractions that differ from $0.5$ leads to more dramatic
effects of the changes in flow parameters. Indeed,   the system properties are   invariant by the transformation~:
 \begin{equation}
   \begin{array}{rcl}
     c &\to & 1-c \\
     \nu_1 &\to & \nu_2\\
      \nu_2 &\to & \nu_1
   \end{array}
 \end{equation}
 This implies that  when changing $\varphi$ to $0.5 +(0.5-\varphi)$ and $VC$
 ($\log VC$) to $1/VC$ ($-\log VC$), the quantities that depend on the
 orientation of the interface  such as the mean average curvature
 $<\kappa_m>$  are transformed into $-<\kappa_m>$ while quantities such as the
 average Gaussian curvature $<\kappa_g>$  that are independent of the interface
 orientation are transformed into  $<\kappa_g>$.   As a result in the parameter space
 ($\varphi,\ \log VC$), ($\varphi=0.5$, $VC=1$) is a center of symmetry that corresponds 
 to  a point where the interface has zero averrage mean curvature and to an extremum of the averrage gaussian curvature }
  As a result  the vicinity of ($\varphi=0.5$, $VC=1$)
 reflects the symmetries of the problem: more specifically, the average
 Gaussian curvature when changing parameters is  marginally affected  which
 implies that the connectivity of the bicontinuous structure is mostly
 unchanged.  In the following we will show  that in the more general case where 
 $\varphi\neq0.5$ this is no longer the case in the self similar regime. In
 addition we will present a description of the loss of stability of the self
 similar-regime and give a rationale for the transition  inspired by \cite{Bouttes2016}.

\subsection{The self similar regime}

 \begin{figure} \centerline{\includegraphics[width=0.6\textwidth]{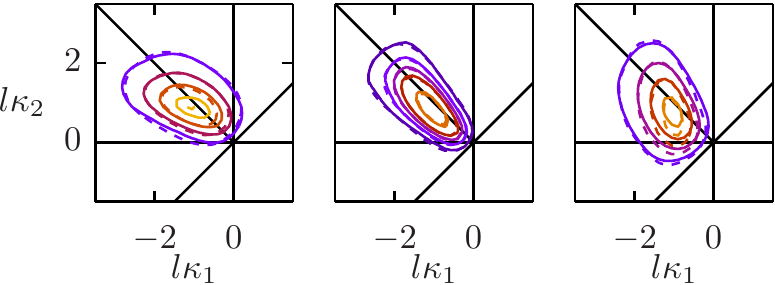}}

 \centerline{\includegraphics[width=0.6\textwidth]{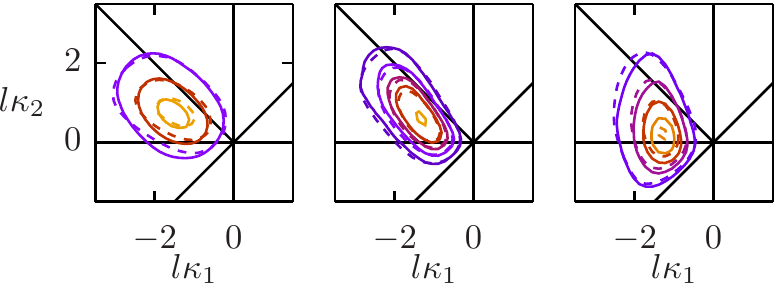}}

   \caption{contour plot of the PDFs of the principal curvatures for
   $\varphi=0.45$ (top) and   $\varphi=0.35$ (bottom) and $VC$  equal to 1/128 , 1
   , and 128 from left to right. The sign convention is chosen so that the
   negative curvatures correspond to center of curvature in the minority phase
   (normal vector pointing toward the minority phase).
   One can see that there is a clear change in the shape of the distribution.
   Nevertheless, the self-similar nature of the coarsening regime is illustrated
   by the fact that the rescaled PDFs are independent of the characteristic size
   of the microstructure for two different values of the characteristic length.
   \label{fig_contours}}
 \end{figure}

 For a wide parameter range around the perfectly symmetric case,  an initially
 bicontinuous structure evolves  after a transient regime in a self-similar
 manner as it has been described previously.  The morphology of the self-similar 
 structure is affected by the control parameters (the volume fraction of the
 phase '1 and the ratio of the viscosities of the phase 1 and of the phase 2).  
 In order to give a clear picture of the effects of varying both parameters, we
 will describe the effects of changing   the relative viscosity of the
 majority phase with respect to the minority phase for a given value of its
 volume fraction. We will also describe the effects of changing the volume
 fraction for a given relative viscosity of the less viscous phase with respect
 to the more viscous one. One should note that while the former approach had
 already been presented in a previous work it was restricted  to variations around
 the symmetric point and it was limited to a volume fraction of 0.5. This, because of  symmetries,  implied that the range over which the ratio of viscosities could be varied was
 limited to two order of magnitudes (between 1 and 128 ). Here since we no longer consider the
 perfectly symmetric case, the range over which the ratio can be varied is 4
 orders of magnitude (between 1/128 and 128). In addition, the parameter value $VC=1$ is no longer  a
 center of symmetry. { Hence, thanks to the departure from the vicinity of the center of symmetry and to the wider range of viscosity contrast that can be explored, more  visible changes of the microstructure induced by tuning the flow parameters are expected. }

 This is well illustrated in figure \ref{snapshots_VC} where, for three
 different values of the viscosity contrast between the phases, the interface
 between the two phases is plotted at a time of a simulation where the
 self-similar regime is established and  for approximately the same
 characteristic lengthscales $l$. The volume fraction of the minority phase is
 $\varphi=0.3$ and the interface is plotted (and coloured proportionally to its
 Gaussian curvature) when it is 4 times more viscous, 4 times less and 16 times
 less viscous than the majority phase. From these picture, it is clear that the
 microstructure are different. When the minority phase is more  viscous, regions 
 of positive Gaussian curvature can hardly be seen on the interface. On the
 contrary, when the minority phase is made less viscous the surface area of  regions
 with positive Gaussian curvature on the interface is increasing. It should also
 be noted that these regions correspond to spherical caps of the minority phase
 protruding  in the majority phase(both $\kappa_{1,2}<0$). 
 Hence when decreasing the relative viscosity   of the
 minority phase, the microstructure is evolving from a structure that is a
 network  of capillary bridges that are close to minimal surfaces with zero
 mean curvature and negative Gaussian curvature  to a similar structure with the
 addition of multiple buds  of the minority phase protruding in the majority
 phase.  
\begin{figure}
  \includegraphics[width=0.6\textwidth]{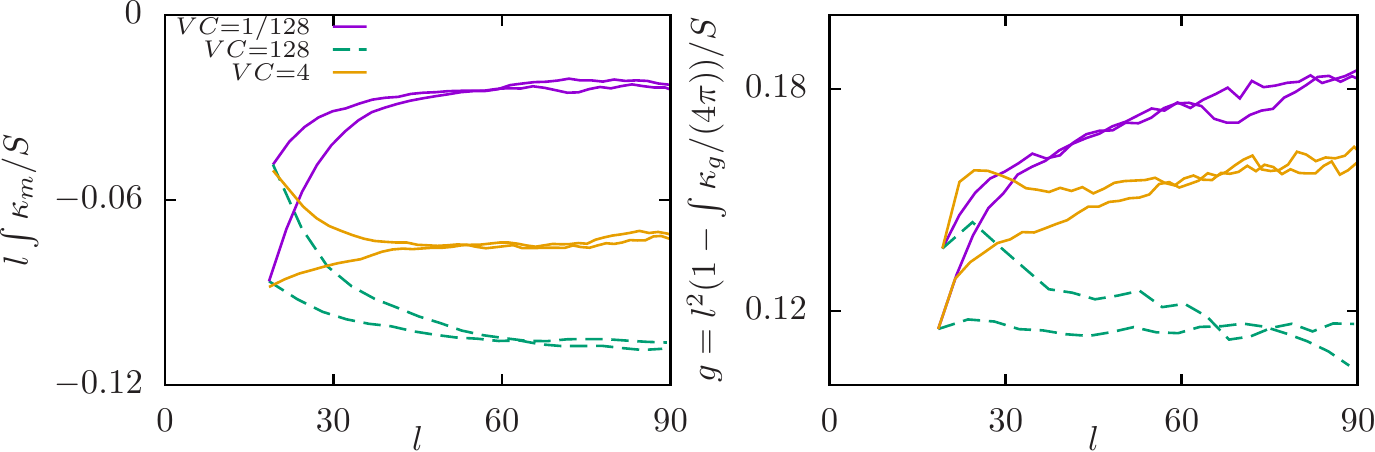}
  \caption{\label{fig_evolmeangauss}Evolution as a function of the
  characteristic length  of the rescaled mean curvature (left) and  genius (right) for the same
  volume fraction $\varphi=0.40$ and different values of the viscosity ratios. For each value of the viscosity ratio two simulations are represented with different initial conditions and different initial values of the genius and average mean curvature. }
\end{figure}

  In the following parts of this section we will present evidences of the
  self-similar nature of the coarsening regime and  quantitative measures of
  the effects of changing the viscosity contrast and the volume fraction on  the
  microstructure. The
  self similar nature of the coarsening regime is well illustrated in
  figure \ref{fig_contours} where the contour of the PDFs of the rescaled principal
  curvatures are plotted for different values of the viscosity contrast and
  volume fraction ($\varphi=0.45$ and $\varphi=0.35$). On each plot contours have been plotted at different
  times corresponding to   $l\approx 40$ and $l\approx 80$ and they superimpose
  well, which indicates that the coarsening process is self-similar as it was
  for $\varphi=0.5$. The effects on the patterns of changing both the volume fraction and the
  viscosity contrast are detailed in the following.
\begin{figure}
  \centerline{\includegraphics[width=0.6\textwidth]{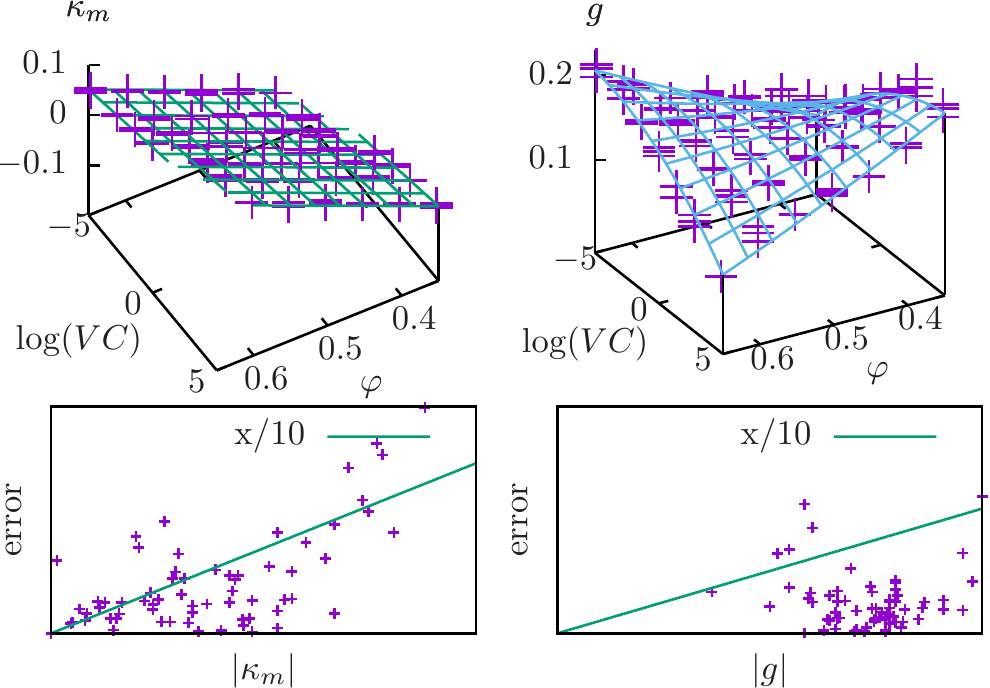}}
  \caption{\label{fig:3dplotcurvs}Plot of the rescaled genius and mean curvature
  as a function of the volume fraction of the phase 1 and the log of the
  viscosity contrast. The surfaces correspond  polynomial fit of the computed
  values.  The bottom plot correspond to the absolute difference between the
  computed values and the interpolating surface that is represented. One can see
  that the relative error is of the order of 1/10 in both cases.}
\end{figure}

  When the volume fraction is  $\varphi=0.45$, close to
  $\varphi=0.5$, the PDFs are very similar to the one presented
  in\cite{henrytegze2017} and recalled in fig.\ref{fig50}.  On each plot  two rescaled PDFs taken for
  $l\approx 40$ and $l\approx 80$ are represented and superimpose well.  For $VC=1$, the PDF is simply slightly shifted away
  from the zero mean curvature  axis $y=-x$ that is a symmetry axis of the PDF for $\varphi=0.5,\
  VC=1$ and for $VC=128,\ 1/128$, they are also not very different. One should
  however, note that in the case $VC=128$, the part of the PDF that corresponds to a
  positive Gaussian curvature corresponds to caps of the minority phase (the
  least viscous)  protruding in the majority phase while for $VC=1/128$, it corresponds to caps
  of the majority phase (the least viscous) protruding in the minority phase.  
\begin{figure}
   \includegraphics[width=0.6\textwidth]{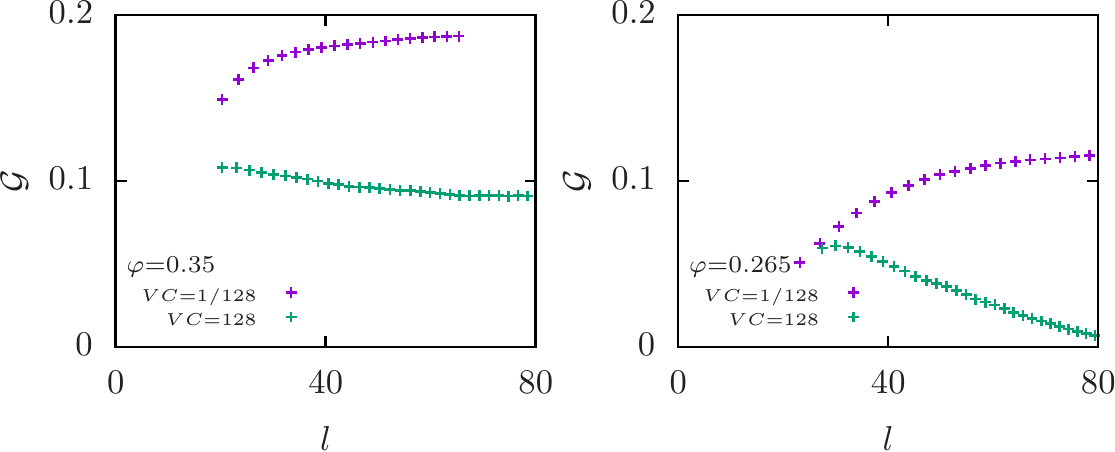}
   \caption{Conductance of the minority phase during the coarsening for two
   values of the viscosity contrast and two values of the volume fraction.
   \label{figconductivities_1}  }
 \end{figure}

  For a smaller volume fraction $\varphi=0.35$, if both phases share the same
  viscosity there is no significant departure from the shape presented for
  $\varphi=0.45$. The shift from  the symmetry axis is simply  more pronounced. When the
  minority phase is significantly less viscous ($VC=128$),  a significant
  part of the interface  corresponds to region of  positive Gaussian
  curvature. Hence, as can be  seen in fig.\ref{snapshots_VC}, an important part
  of the minority phase consists of protrusions in the majority phase that 
  do not participate to the connectivity of the minority phase
  cluster. In  the case where $VC=1/128$, the PDF maximum is close to the
  $\kappa_2=0$ axis that corresponds to cylindrical part of the interface (and
  zero Gaussian curvature). This corresponds to the very thin filament
  that can be seen for instance in  in figure\ref{snapbreak}(d).   
\begin{figure}
   \includegraphics[width=0.6\textwidth]{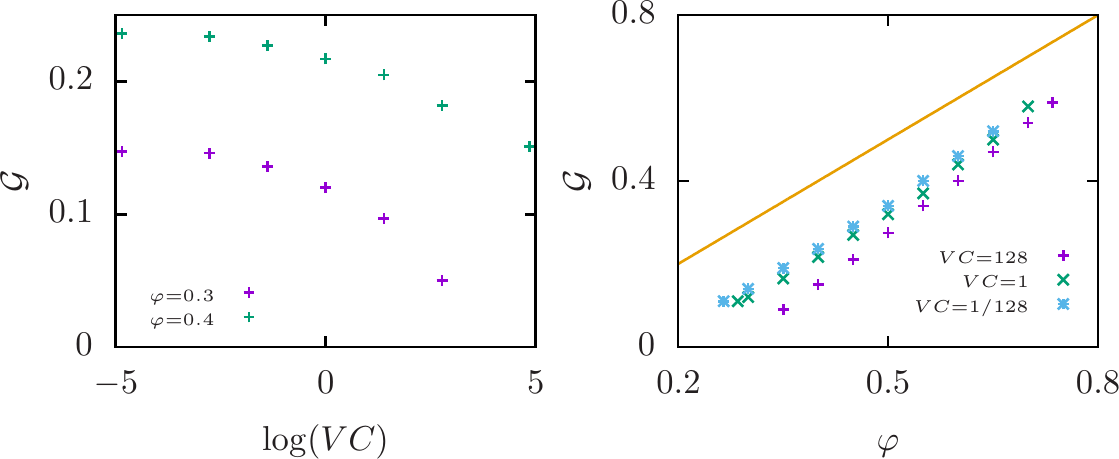}
   \caption{\textbf{Left} Conductance of the sample as a function of the
   logarithm of the viscosity contrast for two values of the volume fraction
   $\varphi$. \textbf{Right}: conductance
    as a function of  the volume fraction when the viscosity of the
   conducting phase is 128, 1 and 1/128 times the viscosity of the isolating
   phase. The straight line corresponds to $\mathcal{G}=\varphi$
   \label{figconductivities_2}}
 \end{figure}

  In figure \ref{fig_evolmeangauss} the evolution of the average rescaled mean and
  Gaussian curvatures are plotted as a function of the characteristic length
  during coarsening for $\varphi=0.40$, different flow conditions (each line
  type corresponds to a given flow condition) and different initial conditions
  (this is made clear by the fact that the lines start at different values of
  $l\kappa_m$ and $g$). In both cases,  after a transient that
  corresponds roughly to doubling $l$ a stationary regime is reached and 
  the limit values of the average Gaussian and mean curvatures are functions  of the flow
  parameters and volume fraction and are independent of the initial condition\footnote{It should be noted that the convergence of the Gaussian curvature toward its limiting value is slower.}. 
   This convergence allows to compute the rescaled
  mean curvature and genius as functions of the
  viscosity contrast and the volume fraction. The results are summarized in 
  fig. \ref{fig:3dplotcurvs}, for the range of parameters considered here. 
  The mean curvature is very well approximated by a
  plane. The rescaled genius plot shows that the center of symmetry
  $\varphi=0.5,\ VC=1$ is not an extremum but a saddle point. However, despite
  the fact that the genius is a topological invariant, interpreting this plot in
  terms of connectivity of the microstructure is difficult. 
  
   In order to actually quantify the effects of the control parameters on
  the connectivity, we have computed the electrical conductance of the microstructure
  assuming that one phase is conducting (with conductivity 1) while the other is
  isolating (with conductivity $\approx 0$), the values of the domain size are chosen so that the conductance of
  a sample filled with the conducting phase is 1 (details are given in the
  appendix). In fig. \ref{figconductivities_1} (a) the evolution of
  the conductance of the microstructure as a function of $l$ is plotted  for two values of
  the viscosity contrast and a value of the volume fraction of the conducting
  phase $\varphi=0.35$. The plot indicates that the conductance is converging
  toward a limiting value during the coarsening and that the higher
  conductance is reached by the more viscous phase.

  This is confirmed by the plot of the conductance as a function of the viscosity
  contrast $VC=\nu_/1/\nu_2$ where the phase 1 is conducting while the phase 2 is
  isolating for $\varphi=0.3$ and  $\varphi=0.4$ in  fig.
  \ref{figconductivities_2}. One can see that when decreasing the contrast of
  the minority phase (which is conducting), there is first a region for which
  the conductance is not changing a lot while the morphology is changing as
  is seen on the PDFs of the curvatures or on the evolution of the average  mean
  curvature. Then when the viscosity is decreased further, the conductance
  decreases significantly. Since the conductance of the phase 1 cannot be larger
  than its volume fraction (with the conventions used here), the existence of
  the aforementioned plateau is  obvious. Our simulations did not allow
  to explore systematically the effects of varying the volume fraction on the
  position of the threshold.   
  
  Finally in fig.
  \ref{figconductivities_2} we have plotted the conductance $\mathcal{G}$ as a function
  of $\varphi$  for different values
  of the viscosity contrast. It is clear that, for a given
  value of $VC$, there is a threshold of the volume fraction below which the
  conductance of the minority phase goes to zero. This implies that the
  microstructure is no longer bicontinuous and that the coarsening regime is no
  longer a self similar viscous regime.  The computed conductances when
  approaching this transition point are decreasing linearly a function of the
  control parameter (either $\varphi$ or $\log(VC)$ ) down to very small values
  of the conductance.
  This indicate that
  it is likely that the transition from the bicontinuous phase to the inclusion
  in a matrix phase is continuous: there is no threshold parameter (e.g.
  $\varphi$ for a given $VC$) above which there exist a bicontinuous phase with
  a finite conductance  and below which the conductance goes to zero.
   { As a result  the limits of the self-similar
  regime should correspond to the domain where $\mathcal{G}(\Phi,VC)=0$  which can be
  extrapolated using the curves in fig. \ref{figconductivities_2}. However, such  an   extrapolation is likely to give unphysical results when  
  $\varphi$ is close to 1 or to 0 and must be taken with care.}

\begin{figure} 
  \includegraphics[width=0.6\textwidth]{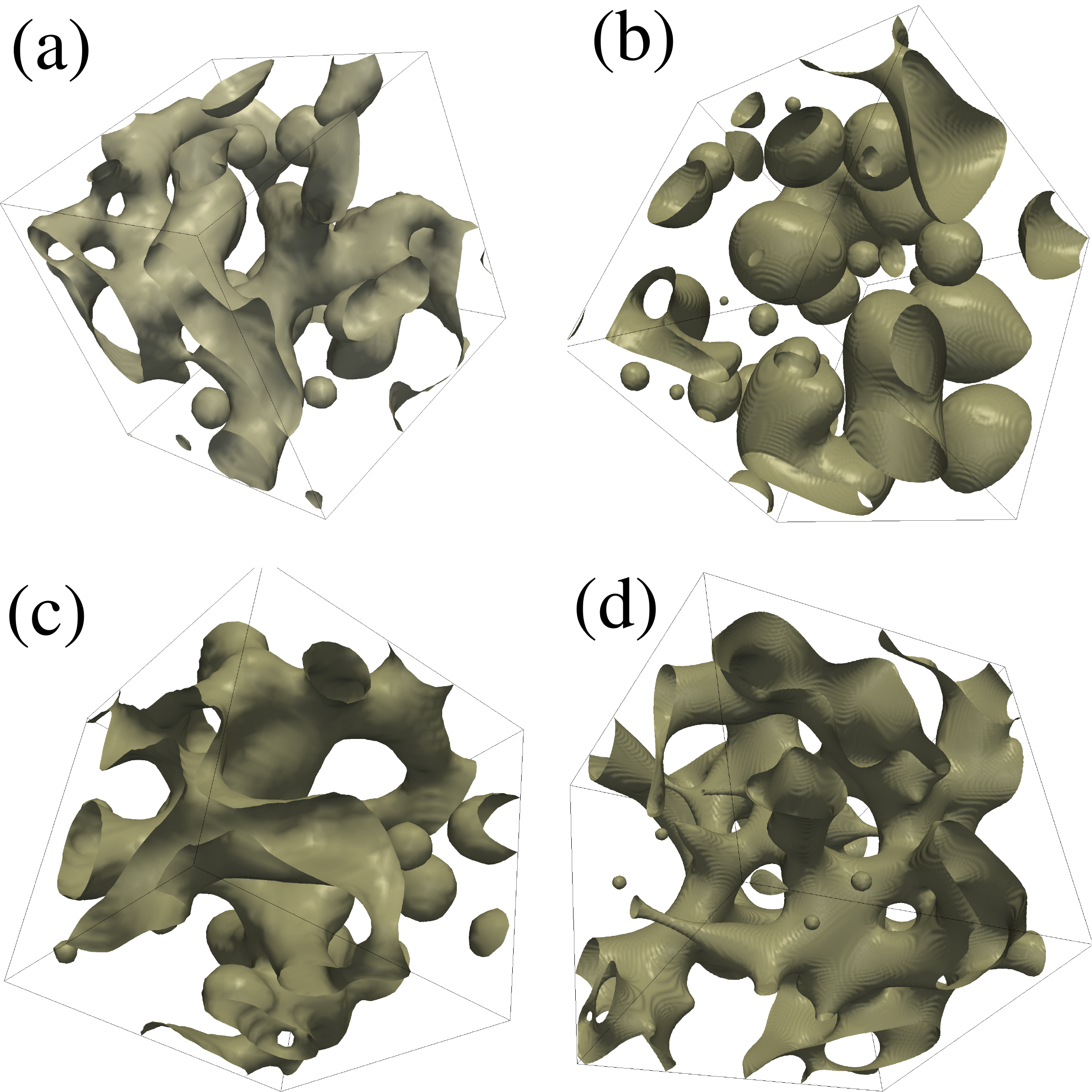}

  \caption{(a),(b) snapshot of the microstructure corresponding to (a) and (d)
  in fig. \ref{figPDFbreakup}. (c) and (d) correspond to (e) and (h) in  fig.
  \ref{figPDFbreakup}. In all snapshots a portion of the domain  of size proportional to
  the characteristic lengthscale $l$ is shown.\label{snapbreak}}
\end{figure}

\begin{figure}
   \includegraphics[width=0.6\textwidth]{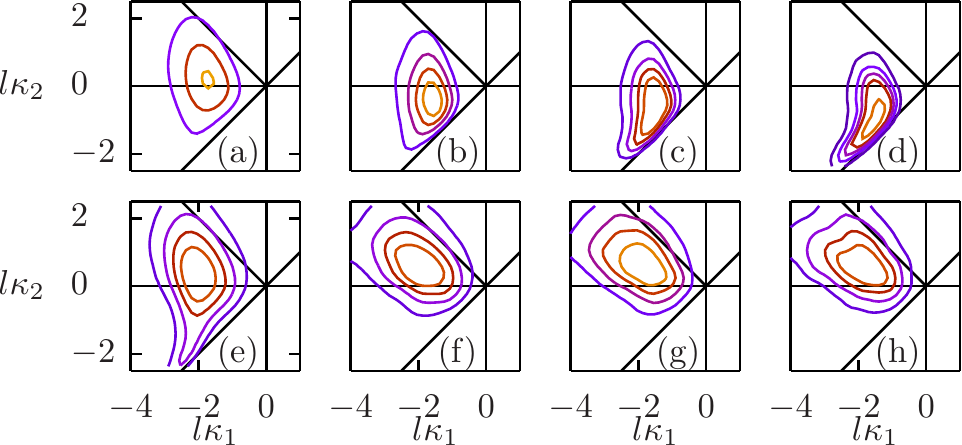}

   \caption{a, b, c and d: Evolution of the PDFs of curvatures for
   $\varphi=0.735$ when the minority phase is 128 times less  viscous than the
   majority phase.  The PDFs is initially centered relatively close to the
   $\kappa_1=-\kappa_2$ line. The evolution is such that it is moving toward the
   line $\kappa_1=\kappa_2<0$ that corresponds to inclusions of the minority phase
   in the majority phase. e, f, g and h: the same evolution when the minority
   phase is the more viscous one. There is a clear evolution toward a
   distribution centered close to the  $\kappa_1=-\kappa_2$ line (that
   corresponds to $\kappa_m=0$ that appears to be almost unaffected by the changes in
   $l$.\label{figPDFbreakup}}
 \end{figure}

\begin{figure}
  \includegraphics[width=0.6\textwidth]{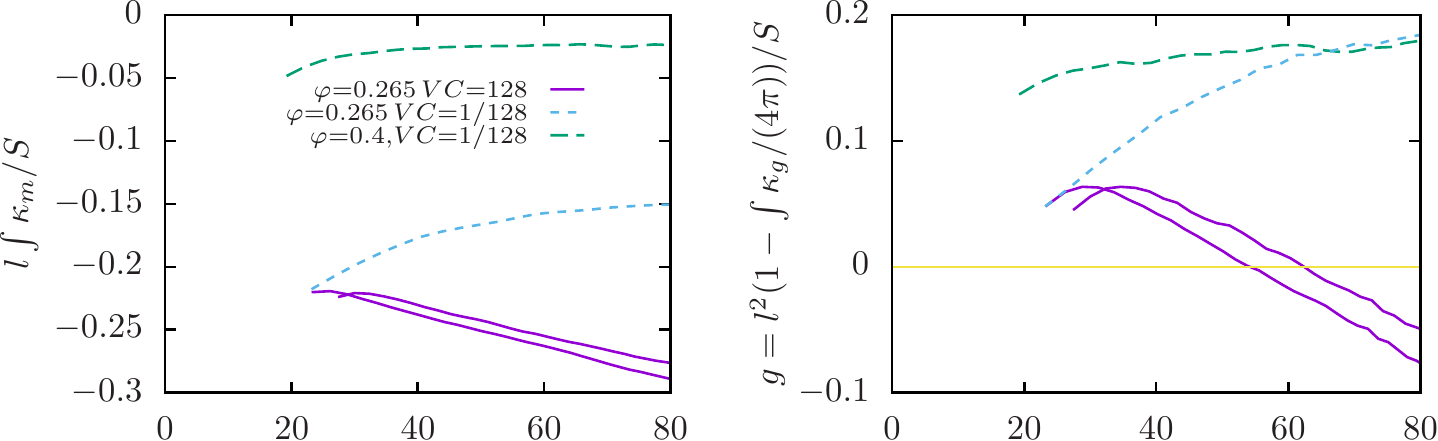}
  \caption{\label{figevolmean735}Evolution with characteristic length  of the
  mean curvature (left) and rescaled genius (right)for the same volume fraction
  $\varphi=0.265$ (and  $\varphi=0.4,\ VC=1/128$ as a visual reference)  and different values of the viscosity ratios. The values
  interpolated from the fit of the self-similar regime close to the perfectly
  symmetric point  are: $\kappa_m= -0.18,\ -0.10$, $g=0.20,\ 0.06$ when $VC=128,
  1/128$.}
\end{figure}

  Now we give a short
  description of the transition from a bicontinuous microstructure toward an
  inclusions in matrix pattern. To this purpose we consider the evolution of
  an initially bicontinuous pattern when the volume fraction of the minority
  phase is 0.265. For this value of the volume fraction, there is a self-similar
  coarsening regime when the minority phase is 128 times more viscous than the
  majority phase as can be seen in fig \ref{figPDFbreakup} and \ref{snapbreak}. Starting from a microstructure obtained in this regime we let
  the system evolve with a minority phase that is 128 times less viscous than
  the majority phase. Snapshots of  the microstructure and of
  the PDFs of the principal curvatures during its evolution are represented in
  in fig \ref{figPDFbreakup} and \ref{snapbreak} . For the sake of readability, the snapshots of the microstructure 
  have been taken using portions of the simulation domain of varying size proportionally 
  to the characteristic length $l$. Both the snapshots and the PDFs show that
  starting from a bicontinuous pattern that consists of capillary bridges linked
  in a network, the microstructure evolves toward a pattern where there are
  less capillary bridges and more spherical caps on the microstructure. This
  evolution eventually leads to the formation of multiple inclusions of the
  minority phase isolated in the majority phase.  In fig.
  \ref{figconductivities_1}  the changes
  in the conductance of the microstructure during this evolution are  plotted
  and  there is a linear decrease of $G$ with $L$ until $G$ reaches zero, which
  indicates that the minority phase is no longer percolating.  It should also be
  noted that during the time evolution from a bicontinous structure to an incluions in a mattrix pattern, 
  neither the averrage mean curvature nor the the averrage gaussian curvature (see fig. \ref{figevolmean735}) present a discontinuity. 
  They vary smoothly: when the self similar regime is unstable, the mean
  curvature and the gaussian curvature decrease linearly with time. When
  considering the two curves it is impossible to detect the loss of continuity
  of the microstructure. Hence, since the genius is a topological invariant, this 
  confirms that the evolution of the network is continuous in time in the  large system size limit.

{{
 These results on both the self similar regime and the loss of self-similarity
 can be understood using an argument inspired by the work of
 \cite{Bouttes2016}. Indeed, the coarsenning of a bicontinuous  microstructure  
 consists of the successive breakup of capillary bridges (of the minority phase
 if $\phi$ differs significantly from from 0.5).
 Once a capillary
 bridge is  broken it retracts and the fluid that was present in it is
 spead  in the remaining  capillary bridges, making them thicker and
 therefore delaying  their break-up. Such a process is possible only if
 the retraction time of the filament is small  enough when compared to the breakup
 time of the remaining filaments. 
 
  In \cite{Bouttes2016}, the authors show that the retraction time of a filament of a fluid of viscosity $\nu_{fil}$
 in a composite matrix consisting of the same fluid as the filament and a
 complementary phase with viscosity $\nu_{comp}$ is a linear combination of 
 $\nu_{comp}$ and $\nu_{fil}$. They also show that, when there is a strong viscosity contrast between the phases, the filament
 breakup time is proportional to $\sqrt{\nu_{comp}\nu_{fil}}$  when
 $nu_{comp}>>\nu_{fil}$ and to $\nu_{fil}$ when
 $nu_{fil}>>\nu_{comp}$. As a result, the retraction time of a viscous filament
 is comparable to its breakup time while the retraction time of a fluid filament
 is much larger than its breakup time\footnote{These
 charactetristic time are also function of the geometric characteristics of the
 filament. However, in the case of self similar coarsenning the effects of the
 geometry are similar  for both characteristic times.}.

 This allows us, for instance to
 give a rationale to the increased presence  of spherical caps in the
 microstructure when the minority phase is made less viscous than the majority
 phase. Indeed, the retraction time of a broken filament (with spherical caps)  is inversely
 proportional to the rate of disapearence of these spherical caps while the
 capillary breakup time is inversely proportional to the rate of appearance of
 these broken filaments (spherical caps). Therefore, when the viscosity of the
 minority phase is increased, the number  of spherical caps in the
 microstructure which is
 the ratio of these two rates is increased (at dominant order in the limit of large systems). 

 This also gives a good understanding of the loss of self-similarity when the
 viscosity of the minority phase is decreased. Indeed, the self similar regime
 is possible if there is a balance between the flux of mass that goes into the
 broken
 filament due to capillary breakup and the flux of mass that goes from the
 broken filaments to the network of unbroken  filaments of the minority phase
 that make the minority phase continuous. These two fluxes are respectively
 inversely proportional to the capillary breakup time and the retraction time.
 If the later is not large enough, such a balance cannot be achieved by the
 system and loss of self-similarity is observed. With such a process, one would
 expect to observe a progressive decrease of the conductivity of the pattern as
 can be seen in fig. \ref{figconductivities_1}.
 
 Hence the effects of the changes in the flow parameters on the microstructure are  related to their effects on the  capillary bridge break-up and capillary bridge retraction characteristic times. 

 }}

\section{Conclusion}
We have studied the evolution of the microstructure of a biphasic fluid
under the action of surface tension and have focused our work on the effects of
considering volume fractions that differ significantly from 0.5 and 
fluids with different viscosities. From our simulations, it appears that the self
similar regime is robust to departure from the perfectly symmetric point for
which it had already been observed. It must be noted that for a wide range of
parameters and a wide range of initial conditions, it is  an attractor.
However when the volume fraction differs significantly from 0.5, the initially
bicontinuous microstructure  evolves under the action of flow in a non self
similar manner and eventually become a set of  inclusions in a matrix. A
few example of this transition were numerically studied and in all cases, the
transition was not accompanied by abrupt transition of quantitative observables
such as the average mean or Gaussian curvature or the conductance of one phase. 
The fact that when the minority
phase is less viscous than the majority phase, this transition is favoured can
be interpreted in the light of the direct observation of the morphological
characteristics of the microstructure. It is   a consequence of a relative
increase of the retraction time of liquid filament after the breakup   of
capillary bridges when compared to the characteristic time for breakup. This
is in line with the mechanism  was initially proposed by
Bouttes\cite{Bouttes2016} in the light of
experiments: the loss of stability of the self similar coarsening regime when
the viscosity of the minority phase is decreased is due to the fact that the
time for retraction of a filament is becoming much larger than the
characteristic time for capillary bridges. 

 More importantly, we have shown that the kinetics of coarsening have a
 dramatic effect on the microstructure and in the case of viscous coarsening we
 have been able to show their importance In addition, our results indicate that
 the perfectly symmetric regime  (where exchanging  phases does not change the
 problem) is a very peculiar point due to symmetries,  and that considering more general situations
 gives more insight on the pattern forming process.

\begin{acknowledgments}
The authors would like to thank D. Vandembrouq, E. Gouillart and K. Thornton for stimulating
discussions. Travel expenses that permitted the cooperation were covered by the
PICS program from CNRS and computations were performed at IDRIS under the
allocation A0042B07727.
\end{acknowledgments}

\appendix
\section{\label{appendixmono}Rescaled mean curvature and genius of a monodisperse suspension of inclusions}
In the main part of the manuscript the genius and average mean curvature are
used extensively and they are supposed to measure to some extent the morphology
of the microstructure. 
While for the complex microtructures presented here they come from the
integration of the curvatures over complex surfaces,  in the case of a
monodisperse set of inclusions in a mattrix, they are
directly related to the volume fraction of one phase  and can be easily computed. Here we
briefly give their values after recalling the steps leading to the values. To
this purpose we consider a monodisperse suspension of spheres that fill the
space with a volume fraction $\Phi$. This situation is reached, for instance
when space is filled with cubes of side 1, that contain, each, a sphere of
radius $r$. The radius of the sphere is then such that:
\begin{equation}
  \frac{4}{3}\pi r^3=\Phi
\end{equation}
which implies 
\begin{equation}
  r=\left(\frac{3\Phi}{4\pi} \right)^{1/3}
\end{equation}
Using the definition of the characteristic length used here, we have that: 
\begin{equation}
  l=\frac{1}{4\pi r^2}=\left(36 \pi\Phi^2 \right)^{-1/3}
\end{equation}
The rescaled average mean  curvature is then (up to sign change depending on the convention)
\begin{equation}
l <\kappa_m > =  l \frac{2}{r}= \frac{2}{3\Phi}
\end{equation}
where $\Phi$ is the volume fraction occupied by the spheres.
For the genius, the same reasonning applies and gives
\begin{equation}
  g=(1-N)
\end{equation}
where N is the number of elementary cubes. From this, the rescaled genius is, in the limit of large $N$
\begin{equation}
  gl^3/V=\frac{1-N}{N}l^3\approx - \frac{1}{ 36\pi\Phi^2}
\end{equation}
For the values used here we have typical values of the rescaled genius and mean curvature that are summarized in the table\ref{table_genius}
\begin{table}
  \begin{tabular}{|c|c|c|c|c|c|c|}
    \hline
    $\Phi$ & 0.25 & 0.3 & 0.35 &0.4 & 0.45&0.5\\
    \hline
    $l<\kappa_m>$ &-2.7&-2.2 &-1.9&-1.7&-1.5&-1.3\\
    \hline
    $-gl^3/V$ &0.14 &0.1 &0.07&0.055&0.044&0.035\\
    \hline
  \end{tabular}
  \caption{\label{table_genius} summary of the rescaled mean curvature and
  genius for different values of the volume fraction, in the case of a
  monodisperse suspension of spheres. Higher values of $\Phi>0.523$ are not considered
  since they would correspond to overlapping spheres.}
\end{table}

\section{Measure of the conductivity of the system}
In order to measure the connectivity of the microstructure (i.e.) of one of the
phases, one can measure the conductivity of the microstructure with one phases
with a conductance of 1 and the other a conductance of $g<<1$. This computation,
with a sufficiently small $g$,  will give a measure of the section of the
continuous paths of the conducting phase that go through the sample and
therefore of the connectivity of the sample. This is what has been measured here
by solving the linear PDE for a given microstructure (the size of the
microstrure was set to 1 in all three directions) $c(\mathbf{x})$: 
\begin{equation}
0=\nabla (G(c) \nabla V)\label{eq_laplace}
\end{equation}
whith  $G(c)=1$ if $c(\mbox{resp }(1-c)))>0.5$ and $G(c)=g$ if $c(\mbox{resp
}(1-c)))<0.5$ in order to suppress the possible effects of  the exponential
tails of the interfaces. The boundary conditions are $V(i=L=1)=1 \mbox{ and }
V(i=0)=0$ and periodic at the other faces of the sample where $i$ is either $x$
or $y$ or $z$.  Once $V$ is computed (using a method described below), the flux
along the $i$ direction is:
\begin{equation}
  \Phi_i=\int_{j}\int_{k} g\partial_i V \label{eq_conductance}
\end{equation}
where $j$ and $k$ are the two remaining indices once $i$ is set. The results as
expected from the isotropy of the sample are independent of the choise of $i$.
Therefore the  average value of the flux $\Phi=(\Sigma_i\Phi_i)/3$ can de used
to define the conductance  $\mathcal{G}=\Phi$
 In such a system,  one can easily see that (conducting tubes along the gradient
 of $V$ which is anisotropic) the  maximal conductance (conducting tubes along
 the gradient of $V$ which is anisotropic)   is equal to the volume fraction of
 the conducting phase $\varphi$. 

 When computing V and the conductance of the microstructure,  the description
 of the interfaces is of little interest, therefore  a small
 undersampling, that is using 1 point out of 4 in each direction, was used
 (using 1 point out of 8 did not affect results in all test cases considered). As a result
 the system size used when solving  the discrete version of eq. \ref{eq_laplace}
 was $256^3$ which is large.

 The solution was computed using a discretized damped wave equation with a
 properly chosen damping $\lambda=0.005$ and varying mass density  to ensure
 fast convergence toward the equilibrium and a constant wave equation in domains
 independently of the phase:
 \begin{equation}
   \partial_{tt} V=\frac{1}{G(c)}(\nabla (G(c) \nabla V))-\lambda\partial_t V
 \end{equation}
 Simulations showed that after $10^4$ iteration, a very good convergence had
 been reached~: the residual were extremely small and the value of $\Phi$ that
 was computed was nearly independent of the position where it is computed. This
 was in stark contrast with results obtained using Gauss Seidel over-relaxation
 for which after a comparable number of iterations, the same value of the
 error(using $L_\infty$ norm) was reached  but where significant  long wavelength variations
 of the flux $\varphi$ were present. An estimate of $\lambda$ in the case of a
 continuous one dimensional system is of the order of magnitude of  $ c/L$ where
 c is the wave-speed and $L$ is the size of the system: considering higher
 values of $\lambda$ would lead to a mode whose amplitude decreases with a rate
 much lower than $\lambda$. 

 The algorithm, which is straightforward, was implemented using GPU acceleration
 and double precision and   solutions of one given problem of dimension $256^3$
 were  reached within approximately 40s. (Using a NVIDIA Tesla P100 Card).   
 \section{Initial condition}
 The initial condition were computed with the following two step algorithm:
 \paragraph{First} the computation domain was filled with oblates ellipsoids
 (that could overlap) of one phase until the desired mean concentration of one
 phase was reached. The choice of prolate ellipsoids  allows to reach low volume
 fraction while keeping a bicontinuous structure.
 \paragraph{Second} the system was evolved for a relatively short time that
 corresponds to a significant increase in l (typically by a factor of 2) with
 different kind of kinetics: either purely diffusive or with a Navier Stokes flow
 and a viscosity contrast that was (or was not) the one  to be
 used in the main run.

 A given initial condition was used for different simulations using different
 parameter values for the flow.
\end{document}